\documentclass[manuscript]{aastex}

\def \ergsec{\hbox{erg$\,$s$^{-1}$}}

\def \gray {$\gamma$-ray }

\def\cp{3C 454.3~}
\def\c{3C 454.3~}

\def\rffa{}

\usepackage{epsfig}

\shorttitle{The MWL campaign of Dec. 2009  on 3C 454.3}

\begin{document}
%
%\title[The mwl campaign of Dec. 2009  on 3C 454.3]{A comprehensive high energy view of the gamma-ray flares of 3C 454.3 in December 2009}

%aastex emulateapj formatting
%\title[The MWL campaign of Dec. 2009  on 3C 454.3]{The December 2009 gamma-ray flare of 3C 454.3: the multifrequency campaign}

%aastex formatting
\title{The December 2009 gamma-ray flare of 3C 454.3: the multifrequency campaign}

\author{L. Pacciani\altaffilmark{1}, V. Vittorini\altaffilmark{1,2}, M. Tavani\altaffilmark{1,2}, M. T. Fiocchi\altaffilmark{1},
 S. Vercellone\altaffilmark{17}, F. D'Ammando\altaffilmark{17}, T. Sakamoto
\altaffilmark{22,27,28},
  E. Pian\altaffilmark{16,23}, C. M. Raiteri\altaffilmark{24},
M. Villata\altaffilmark{24}, M. Sasada\altaffilmark{37}, R. Itoh\altaffilmark{37},
M.~Yamanaka\altaffilmark{37,38}, M. Uemura\altaffilmark{38}, E. Striani\altaffilmark{2,11}, S.
 D. Fugazza\altaffilmark{25}, A.
Tiengo\altaffilmark{3},  H. A. Krimm\altaffilmark{22,28,29}, M. C. Stroh\altaffilmark{30}, A.
D. Falcone\altaffilmark{30}, P. A. Curran\altaffilmark{21}, A. C. Sadun\altaffilmark{26},
A. Lahteenmaki\altaffilmark{31}, M. Tornikoski\altaffilmark{31}, H. D. Aller
\altaffilmark{32}, M. F. Aller\altaffilmark{32}, C. S. Lin\altaffilmark{33}, V. M. Larionov
\altaffilmark{34,35,36}, P. Leto\altaffilmark{39}, L. O. Takalo\altaffilmark{40}, A. Berdyugin
\altaffilmark{40}, M. A. Gurwell\altaffilmark{41}, A. Bulgarelli\altaffilmark{5}, A. W. Chen
\altaffilmark{3,4}, I. Donnarumma\altaffilmark{1}, A. Giuliani\altaffilmark{3}, F. Longo
\altaffilmark{6}, G. Pucella\altaffilmark{13}, A. Argan\altaffilmark{1}, G. Barbiellini
\altaffilmark{6},
% F. Boffelli\altaffilmark{7},
P. Caraveo\altaffilmark{3}, P. W. Cattaneo
\altaffilmark{7},
% V. Cocco\altaffilmark{1},
E. Costa\altaffilmark{1}, G. De Paris\altaffilmark{1}, E.
Del Monte\altaffilmark{1}, G. Di Cocco\altaffilmark{5}, Y. Evangelista\altaffilmark{1}, A.
Ferrari\altaffilmark{18}, M. Feroci\altaffilmark{1}, M. Fiorini\altaffilmark{3},
% T. Froysland\altaffilmark{2,4},
F. Fuschino\altaffilmark{5}, M. Galli\altaffilmark{8}, F. Gianotti\altaffilmark{5},
C. Labanti\altaffilmark{5}, I. Lapshov\altaffilmark{1}, F. Lazzarotto\altaffilmark{1}, P.
Lipari\altaffilmark{9}, M. Marisaldi\altaffilmark{5},
% M. Mastropietro\altaffilmark{10},
S. Mereghetti\altaffilmark{3}, E. Morelli\altaffilmark{5}, E. Moretti\altaffilmark{6}, A.
Morselli\altaffilmark{11}, A. Pellizzoni\altaffilmark{20}, F. Perotti\altaffilmark{3}, G.
Piano\altaffilmark{1,2,11}, P. Picozza\altaffilmark{2,11}, M. Pilia\altaffilmark{12,20},
% G. Porrovecchio\altaffilmark{1},
M. Prest\altaffilmark{12}, M. Rapisarda\altaffilmark{13}, A.
Rappoldi\altaffilmark{7}, A. Rubini\altaffilmark{1}, S. Sabatini\altaffilmark{1}, P. Soffitta
\altaffilmark{1}, M. Trifoglio\altaffilmark{5}, A. Trois\altaffilmark{1}, E. Vallazza\altaffilmark{6},
% A. Zambra\altaffilmark{3},
D. Zanello\altaffilmark{9},
% L. A. Antonelli\altaffilmark{19},
S. Colafrancesco\altaffilmark{14},
C. Pittori\altaffilmark{14}, F. Verrecchia\altaffilmark{14}, P. Santolamazza\altaffilmark{14},
F. Lucarelli\altaffilmark{14}, P. Giommi\altaffilmark{14} and L. Salotti\altaffilmark{15}}
\affil{$^{1}$INAF/IASF-Roma, I-00133 Roma, Italy}
\affil{$^{2}$Dip. di Fisica, Univ. Tor Vergata, I-00133 Roma, Italy}
\affil{$^{3}$INAF/IASF-Milano, I-20133 Milano, Italy}
\affil{$^{4}$CIFS-Torino, I-10133 Torino, Italy}
\affil{$^{5}$INAF/IASF-Bologna, I-40129 Bologna, Italy}
\affil{$^{6}$Dip. Fisica and INFN Trieste, I-34127 Trieste, Italy}
\affil{$^{7}$INFN-Pavia, I-27100 Pavia, Italy}
\affil{$^{8}$ENEA-Bologna, I-40129 Bologna, Italy}
\affil{$^{9}$INFN-Roma La Sapienza, I-00185 Roma, Italy}
\affil{$^{10}$CNR-IMIP, Roma, Italy}
\affil{$^{11}$INFN Roma Tor Vergata, I-00133 Roma, Italy}
\affil{$^{12}$Dip. di Fisica, Univ. Dell'Insubria, Via Valleggio 11, I-22100 Como, Italy}
\affil{$^{13}$ENEA Frascati,  I-00044 Frascati (Roma), Italy}
\affil{$^{14}$ASI Science Data Center, I-00044 Frascati (Roma), Italy}
\affil{$^{15}$Agenzia Spaziale Italiana, I-00198 Roma, Italy}
\affil{$^{16}$INAF-OATS, Via Tiepolo 11, I-34143 Trieste, Italy}
\affil{$^{17}$INAF-IASF Palermo, Via Ugo La Malfa 153, I-90146 Palermo, Italy}
\affil{$^{18}$Dip. Fisica, Universit\'a di Torino, Torino, Italy}
\affil{$^{19}$INAF-OAR, Monte Porzio Catone, Italy}
\affil{$^{20}$INAF-OAC, localita' Poggio dei Pini, strada 54, I-09012 Capoterra, Italy}
\affil{$^{21}$Mullard Space Science Laboratory, University College London, Holmbury St. Mary, Dorking, RH5 6NT, UK}
\affil{$^{22}$Center for Research and Exploration in Space Science and Technology (CRESST), NASA Goddard Space Flight Center, Greenbelt, MD 20771}
\affil{$^{23}$SNS, Piazza dei Cavalieri, 7, I-56126 Pisa, Italy}
\affil{$^{24}$OATO-INAF, Strada Osservatorio 20, I-10025, Pino Torinese (To), Italy}
\affil{$^{25}$OAB-INAF, via Brera 28, I-20121 Milano, Italy}
\affil{$^{26}$Univ. of Colorado, Denver U.S.A.}
\affil{$^{27}$Joint Center for Astrophysics, University of Maryland, Baltimore County, 1000 Hilltop Circle, Baltimore, MD 21250}
\affil{$^{28}$NASA Goddard Space Flight Center, Greenbelt, MD 20771}
\affil{$^{29}$Universities Space Research Association, 10211 Wincopin Circle, Suite 500, Columbia, MD 21044-3432}
\affil{$^{30}$Department of Astronomy \& Astrophysics, Pennsylvania State University, University Park, PA 16802}
\affil{$^{31}$Aalto University Mets\"ahovi Radio Observatory, Metsahovintie 114, FIN-02540 Kylmala, Finland}
\affil{$^{32}$Dept. of Astronomy, University of Michigan, USA}
\affil{$^{33}$Institute of Astronomy, National Central University, Chung-Li, Taiwan 32054, ROC}
\affil{$^{34}$Astron. Inst., St.-Petersburg State Univ., Russia}
\affil{$^{35}$Pulkovo Observatory, St.-Petersburg, Russia}
\affil{$^{36}$Isaac Newton Institute of Chile, St.-Petersburg Branch}
\affil{$^{37}$Department of Physical Science, Hiroshima University, Kagamiyama 1-3-1, Higashi-Hiroshima 739-8526, Japan}
\affil{$^{38}$Hiroshima Astrophysical Science Center, Hiroshima University, Kagamiyama 1-3-1, Higashi-Hiroshima 739-8526, Japan}
\affil{$^{39}$INAF - Osservatorio Astrofisico di Catania, Italy}
\affil{$^{40}$Tuorla Observatory, Department of Physics and Astronomy, University of Turku, FI-21500 Piikkio, Finland}
\affil{$^{41}$Harvard-Smithsonian Center for Astrophysics, Cambridge, MA, USA}
\email{EMAIL: luigi.pacciani@iasf-roma.inaf.it}

\begin{abstract}

During the month of  December, 2009 the
% Flat Spectrum Radio Quasar
blazar 3C~454.3 became the brightest gamma-ray source in the sky,
% for two weeks,
reaching a peak flux $F \sim 2000 \times 10^{-8} $ph cm$^{-2}$
s$^{-1}$ for E $>\ 100$ MeV.
% The source had triggered the attention of the Astronomers a couple of weeks before, in the middle of November 2009,
% showing a flux  in excess of $500\times 10^{-8}$ ph cm$^{-2}$ s$^{-1}$ for E $>\ 100$ MeV.
Starting in November, 2009 intensive multifrequency campaigns
monitored the 3C~454 gamma-ray outburst.
%  by ground based and in orbit observatories.
%GG%We report the results and the interpretations of the data of the multiwavelength campaign during the flaring activity in gamma rays.
Here we report the results of a 2-month campaign involving AGILE,
INTEGRAL, \emph{Swift}/XRT, \emph{Swift}/BAT, RossiXTE for the
high-energy observations, and \emph{Swift}/UVOT, KANATA,
GRT, REM for the near-IR/optical/UV data. The GASP/WEBT provided radio
and additional optical data.
% On timescales of order
% of a few days, we monitored the very prominent gamma-ray
% super-flare of December 2-3, 2009.

We detected a long-term active emission phase lasting $\sim$1
month at all wavelengths: in the gamma-ray band, peak emission was
reached on December 2-3, 2009. Remarkably, this gamma-ray
super-flare was not accompanied by  correspondingly intense
emission in the optical/UV band that reached a level
substantially lower than the previous observations in 2007-2008.
%GG%
%GG%In fact optical, soft X, hard X-ray and gamma ray data have the same rising starting point, but both the short term,
%GG%both the long term behavior show discrepancies at all frequency, with the possible exception of soft x and hard X-ray data.\\
%GG%

The lack of strong simultaneous optical brightening during the
super-flare and the determination of the broad-band spectral
evolution severely constrain the theoretical modelling.
We find that the
pre- and post-flare broad-band  behavior can be explained by a
one-zone model involving SSC plus external Compton emission from
an accretion disk and a broad-line region. However, the spectra of
 the Dec. 2-3,
2009 super-flare and of the secondary peak emission on Dec. 9, 2009
% show {a moderate} spectral hardening in the gamma-ray range that
%GG%does'nt
cannot be satisfactorily modelled by a simple one-zone model. An
additional particle component is most likely active during
these states.
%GG%an indication
% {the signature} of an additional component.
% together with a moderate optical, X-ray, and hard X-ray
% increase. We find that this spectral behavior is unlikely to be
% accounted for by a standard one-zone model, and we present the
% results of a 2-component spectral modelling.

%GG%severely constrain the theoretical modelling.

\end{abstract}

\maketitle

\section{Introduction}

The flat spectrum radio quasar 3C 454.3 (at a redshift $z=0.859$) turns out to be among the
most active blazars emitting a broad-spectrum ranging from radio
to gamma-ray energies.
Blazars are a sub-class of
active galactic nuclei, with the relativistic jet aligned to the
line of sight. Their spectral energy distributions (SED) typically show
a double humped shape, with the low energy
peak lying between radio and X-rays, and the high energy peak in
the GeV-TeV band \cite{padovani1995}.
% Among blazars, flat spectrum radio
% quasars (FSRQ) have the low energy peak in the radio-IR region,
% and the high energy peak at $\sim$1 GeV.\\
% The low energy region of the spectra can be
% associated with the synchrotron emission from the
% relativistic electrons in the jet. The high
% energy region can be explained by inverse Compton
% emission (leptonic models), with seed photons
% coming from an external region (e.g. the
% accretion disk, the dusty torus), eventually
% reprocessed by the broad line region, or the hot
% corona), or from the synchrotron process itself
% (synchrotron self-Compton).
Detailed description of blazar leptonic emission models can be found in
%GG%\cite{maraschi1992,marscher1992,sikora1994}.
Maraschi et al. (1992); Marsher \& Bloom (1992); Sikora et al. (1994).
The observed spectra can also be modelled in
the framework of hadronic models
%\cite[see][]{mucke2001,mucke2003,bottcher2007},
\cite{mucke2001,mucke2003,bottcher2007}.
%\citep[see][]{mucke2001},
% where the key elements of the emission are
% the very high energy protons of the jet.\\
%GG%The accelerated protons produce proton-synchrotron emission, synchrotron through pair production, and gamma-ray emission through the decay of neutral pion produced in the proton interactions.\\

Starting in 2004-2005,  3C 454.3 showed
a long period of optical activity, with variability timescale
ranging from several months to less than one day. In May 2005,
the source reached a peak magnitude  $R \simeq 12$ showing strong
1-day variability \cite{Villata2006}.
% increased of 1 magnitude in less than one day, .
%(increasing of 1 magnitude in less than one day, and reaching a peak R magnitude of 12 in autumn 2007).
%GGG%Five months after the optical peak emission 3C 454.3 was detected
%GGG%with a radio flux of 13 Jy at 37 GHz, and  14 Jy at 43 GHz.
A radio peak was detected 9 months after the
optical peak, with a flux of 22 Jy at 37 GHz, and  20 Jy at 43
GHz \cite{villata2007}.
%The delayed radio emission was explained by the authors as the combined effect of misalignment of radio emitting region with respect to optical, and to time-changing of jet orientation.
The source was then quiescent from the beginning of  2006 until
mid-2007 \citep[with an R magnitude between 15 and 16, ][]{raiteri2008}. Starting in the
second half of 2007 \cite{vercellone2008}, \c has been detected in
a high  gamma-ray state by AGILE \cite{agile}, and, subsequently
also  by  \emph{Fermi}-LAT \cite{atwood2009,abdo2009a}.
% \cite{atwood2009}, showing
Typically, the level of gamma-ray activity (with a flux of
300-600$\times 10^{-8}$ ph cm$^{-2}$ s$^{-1}$ for E $>$ 100 MeV)
has been observed to be correlated with the optical emission.
Relatively large gamma-ray fluxes were detected during the AGILE
{\rffa{observations of 2007 \cite{vercellone2008,vercellone2009a,donnarumma2009a}.}}
%GGG%July, 2007 observations \cite{vercellone2008}, and the
%GGG%November-December, 2007 observations \cite{vercellone2009a,donnarumma2009a}.
During 2008,  the source that was bright at the beginning of the year, started to fade in
optical band \cite{villata2009}. AGILE
%GG%observed \c from May 2008 to August 2008, and from October 2008 to January 2009, detecting
detected the fading in gamma-rays too \cite{vercellone2009b}.
%GG%The Discrete Correlation Analysis (DCF) performed so far between gamma-ray and optical data yields a delay of $-0.4 _{-0.8}^{+0.6}$  (1 sigma errors) of gamma-ray respect to optical for AGILE
%GG%data and $\sim$ 0 for \emph{Fermi} data.\\
%GGG%AGILE spectral data show significant variability above 100 MeV for the low and high emission states.
\textit{Fermi} reported an
averaged spectrum above 200 MeV with a photon index $\alpha \simeq
2.3 $
% \pm 0.03 \, \rm (stat) \pm  0.09 \,(syst.)$,
and a spectral break at $E_c \sim 2.4 $~GeV, obtained in
August, 2008 \cite{abdo2009a}.
%  \pm  0.3  \rm \, (stat.) \pm
% 0.3 \, (sys.)$~GeV

% \textbf{The broad band spectral energy distribution obtained
% during the December 2007 AGILE multiwavelength campaign suggests a
% contribution to the seed photons coming from a hot external source (e.g. the hot corona)}.\\

% In November/December, 2009 the source showed an extraordinary
% gamma-ray activity, reaching a peak flux $F \simeq 2000\times
% 10^{-8}$ ph cm$^{-2}$ s$^{-1}$ for E $>\ 100$ MeV on 2009 December
% 3, and maintaining a flux in excess of $800\times 10^{-8}$ ph
% cm$^{-2}$ s$^{-1}$ for two weeks \cite{striani2009}.
%
\section{The multifrequency campaign}

The intensive monitoring of \c carried out by our group covered
the period of extraordinary gamma-ray activity in
November-December, 2009 \cite{striani2009}.
% started at the end of November, 2009 and
%ended at the end of December, 2009.
The campaign involved AGILE for the gamma-ray band,
\emph{Swift}/BAT, RossiXTE/HEXTE and INTEGRAL/IBIS in the hard
X-ray band, RossiXTE/PCA and \emph{Swift}/XRT in X-rays,
\emph{Swift}/UVOT in the optical and UV bands, the KANATA
observatory and GRT in the optical, REM in the
near infrared and optical. AGILE observed the source every
day in \emph{spinning mode}, scanning about 70\% of the
whole sky every 6 minutes.
INTEGRAL pointed the source in response to a
Target of Opportunity observations (ToO)
%GGG%proposed by our group
\cite{atelvecellone}, and observed it from  2009 December 6 until
2009 December 12. The RossiXTE satellite observed 3C~454.3 on 2009
December 5 and then daily from December 8 until December 17, 2009
for typical integrations of $\sim$3 ks.
\emph{Swift} started to observe \c on 2009 November 27, in
response to a ToO, and pointed at the source every day
(UVOT performed most of the observations with the UV filters).
% (and performing some observation with the optical filters during the campaign).\\
The KANATA 1.5 m telescope performed a long source monitoring in the V-band,
with a time-step of 1 day.
% starting well before the interesting period.\\
The fully automated 14" GRT (Goddard Robotic Telescope) performed
observations in the V and R bands, quasi-simultaneously with
\emph{Swift},  starting on 2009 November 30. REM started the \c
monitoring  on 2009 December 10 in response to a ToO, and observed
the source every day in the VRIJHK filters.
The GLAST-AGILE Support Program \citep[GASP ][]{villata2008,villata2009}
performed an intensive monitoring-campaign of the source in 2009-2010.
We used a sub-sample of their data:
%GG%\footnote{The GASP was born from the Whole
%GG%Earth Blazar Telescope (WEBT; http://www.oato.inaf.it/blazars/webt/) in September 2007.
%GG%Its aim is to provide long-term continuous monitoring in the optical, near-infrared, and
%GG%radio bands, for a number of bright, gamma-loud blazars.}
Optical observations reported in this paper were performed at:
Lulin, New Mexico Skies, Roque de los Muchachos (KVA), and St. Petersburg.
%GGG%Whereas
GASP radio data were taken at
Mauna Kea (SMA, 230 GHz), Noto (43 GHz), Mets\"ahovi (37 GHz) , and UMRAO (4.8, 8.0, and 14.5 GHz).

%P.I.
%GG%The GASP Consortium
%GG%provided optical data from Lulin, New Mexico Skies, Roque de los Muchachos (KVA), and St. Petersburg.
%GG%The GASP observations at radio frequencies were taken at Mauna Kea (SMA, 230 GHz), Noto (43 GHz), Mets\"ahovi (37 GHz) , and UMRAO (4.8, 8.0, and 14.5 GHz).

%original
%GG%The GLAST-AGILE Support Program (GASP; Villata et al. 2008, 2009) was born from the Whole
%GG%Earth Blazar Telescope (WEBT; http://www.oato.inaf.it/blazars/webt/) in September 2007.
%GG%Its aim is to provide long-term continuous monitoring in the optical, near-infrared, and
%GG%radio bands, for a number of bright, gamma-loud blazars.
%GG%The GASP observations reported in this paper were performed at the following optical
%GG%observatories:
%GG%Goddard (GRT), Lulin, New Mexico Skies, Roque de los Muchachos (KVA), and St. Petersburg.
%GG%As for the radio frequencies, the data used were taken at .

%
\section{Data analysis}
% We summarize here the data analysis of the campaign.
%\subsection{AGILE/GRID Gamma-ray data}
%GG%\subsubsection{AGILE/GRID}
AGILE/GRID data were analyzed using the Build-19 software
% of the Analysis Pipeline,
and the response matrix v10 calibrated in the energy range
100-3000 MeV.
%GG%Counts, exposure, and Galactic background maps were created with a binsize of 0.25 $\times$ 0.25  deg$^2$ for photons with energy greater than 100 MeV.
%GG%Only good quality gamma-ray data were selected,
Well reconstructed gamma-ray events were {\rffa{selected}}
%GGG%in the analysis,
using the FM3.119 filter. All the events collected during the passage in the
South-Atlantic Anomaly were rejected. We filtered out the
Earth-albedo, rejecting
%GGG%those
photons coming from a circular
region of radius 85 deg and centered on the Earth.
%GG%To reduce the systematics due to the knowledge of the effective area,
We rejected photons coming from outside the 35 degrees from the
optical axis. Gamma-ray data were analyzed with integrations of 1
or 2 days, depending on the source flux. We used the standard
AGILE Maximum-Likelihood procedure (ALIKE) \citep[see][for the concept definition]{mattox1996}
for each data set.
%GG%Gamma-ray spectra were obtained with the iterative \textmd{pipeSpettroGen}
%GG%procedure, based on the ALIKE algorithm and the weighted Least
%GG%Squares regression, assuming single power-law spectra from 100 MeV to 3 GeV.
%GG%We notice the reader that \emph{Fermi} reveals a spectral break around 2.5 GeV \cite{abdo2009a}, that we are neglecting in our analysis.
The  integration over 5 weeks from  2009-11-18 to 2009-12-23 UTC yields a photon index 1.88 $\pm$ 0.08 (all the errors reported in the paper are at 1$\sigma$, except where stated).
%GG%, obtained simultaneously fitting the \emph{gal} and \emph{iso} parameters of the Maximum Likelihood.
%
%GG%The spectrum integrated on this timescale is reported in figure \ref{fig:grid_spectrum}, with diamonds for the
%GG%data points, and with a continuous line for the power-law fit.
%
%GG%\begin{figure}
%GG%\centerline{\psfig{figure=grid_spectrum.ps,width=80.mm,height=60.mm}}
%GG%\caption{The spectrum obtained with AGILE/grid integrating data from 2009-11-18 0:00 to 2009-12-23 0:00 UTC.}
%GG%\label{fig:grid_spectrum}
%GG%\end{figure}
%
%GG%The obtained \emph{gal} and \emph{iso} for each energy bin were used in the extraction of the spectral index of the sub-integrations.
%GG%At the days of highest gamma-ray activity the spectra are harder: at MJD 55167.7, the photon index is 1.48 $\pm$ 0.26
%GG%(the spectrum is reported in figure \ref{fig:grid_spectrum} with stars for the
%GG%data points, and with dashed line for the power-law fit), and at MJD 55173.7  is -1.40 $\pm$ 0.34.
%GG%The integration of the two weeks ( from 2009-11-18 0:00 to 2009-12-02 0:00 UTC) preceding the first flare gives a photon index of 1.91 $\pm$ 0.16. The integration (from 2009-12-03 06:00 to 2009-12-08 03:30)
%GG% between the two flares gives a photon index of 2.04 $\pm$ 0.15. The integration over 2 weeks after the flare (from 2009-12-09 09:30 to 2009-12-23 0:00) gives a photon index of 1.90 $\pm$ 0.13.
%
%GG%We note that the spectral indices evaluated during the flares deviate from the mean values of 1.5 - 2 standard deviations.

%\subsection{Hard X-ray data}
% \subsection{INTEGRAL}
The INTEGRAL-IBIS \cite{ubertini2003} data were processed using
the OSA software version 8.0. light curves (from 20 to 200
keV) and spectra (from 18 to 200 keV) were extracted for each
individual science window of revolutions 873 and 874.
% We fitted
% the ISGRI spectrum (integrated for the whole observation) with a
% power law model, yielding a photon index 1.69 $\pm$ 0.11, and a%
% flux of (7.06$_{-0.35}^{+0.15})\times$10$^{-11}$ erg cm$^{-2}$
% s$^{-1}$ in the energy band 20 - 40 keV.

% \subsubsection{\emph{Swift}/BAT}
The \textit{Swift}-BAT survey data were obtained
%GG%from HEASARC \emph{Swift's} Browse interface
applying the ``BAT FOV" option.  The data have been processed by
{\tt batsurvey} script available through HEASOFT software package
with a snapshot (single pointing) interval.  To estimate the
background, ten background points around the source in a radius of
50$^{\prime}$ are selected.  The source, ten background points and
the bright hard X-ray sources (for cleaning purpose) are included
in the input catalog of {\tt batsurvey}.
% Source count rate is
% corrected for the effective area (calibrated with Crab
% bservations).
%GG%If the source is in off-axis location of the field of view, the correction has been
%GG%applied based on the Crab observations, to modify the count rate to the on-axis.
% The daily-averaged count rate of the source
% is estimated by averaging the count rates of snapshot observations during a day.
% The error for the source is estimated by calculating the rms of the count rates in ten background points.
The BAT count rate in the 14-195 keV band
%GG% and in the 15-50 keV band
has been converted into the energy flux
assuming a power-law photon index 1.7 as determined from the
INTEGRAL-ISGRI data, see below). In order to match
with the HEXTE range, the BAT hard X-ray flux has been rescaled in
 Fig. \ref{fig:lc_all} to the 20-40 keV band.
%GGG%assuming a power-law spectrum with photon index 1.7.
%
% \subsection{X-ray data}
%
% \subsection{RossiXTE}

\textit{RossiXTE}-PCA \cite{jahoda1996} and HEXTE \cite{Rothschild1998} data were analyzed following
the same procedure described in
%GG%\cite{vercellone2009b}.
Vercellone et al. (2010).
% In order to minimize the
% uncertainties in the instrument calibration and background
% subtraction,
The data analysis was restricted to the PCU2 in the
3-20 keV energy range for the PCA and to the Cluster B in the
18-50 keV range for the HEXTE. The net exposure times were 27.3 ks
for PCA and 7.3 ks for HEXTE.
%GG%The light curves in the 3-20 keV and 20-40 keV energy bands are shown in figure ...
The background subtracted source spectra obtained with both the
instruments were simultaneously fit\footnote{The resulting fit is
not statistically acceptable ($\chi ^2_{\rm red}$=1.53/54 d.o.f.);
however, the addition of a 2\% systematic error to the data, which
is well within the expected uncertainties in the spectral
calibration, is sufficient to make the fit fully acceptable
($\chi^2_{\rm red}$=1.02/54 d.o.f.). } with an absorbed power-law
model, with the photoelectric absorption fixed to
0.134$\times$10$^{22}$ cm$^{-2}$ \cite{Villata2006}. After the
introduction of a 2\% systematic error, the best-fit value for the
photon index is 1.74 $\pm$ 0.01.

% The source flux in the 3-20 keV and 20-40 keV energy ranges are
% 1.1$\times$10$^{-10}$ erg cm$^{-2}$s$^{-1}$ and
% 5.8$\times$10$^{-11}$ erg cm$^{-2}$s$^{-1}$, respectively. The
% spectrum is therefore slightly softer than the average spectra of
% C~454.3 observed by RossiXTE in 2007 and 2008, when the source flux
% as lower (Vercellone et al. 2010). We also note that  BAT hard
% -ray data were obtained  in the 15-50 keV band.

%
% \subsection{\emph{Swift}/XRT}
The \emph{Swift}-XRT data were processed using
%GG%the most recent versions of the standard Swift tools and
the most recent calibration files available. We utilized
\emph{Swift} Software version 3.5, FTOOLS version 6.8, and XSPEC
version 12.5.1n.
%
% Light curves in the 0.3-10 keV band were generated using xrtgrblc
% version 1.4. All of the observations were obtained in windowed
% timing mode. Box regions rotated to match the spacecraft roll
% angle are used to describe the source and background areas. The
% width of the source rectangle is adjusted depending on the current
% source rate. In order to handle cases where the source landed on
% bad CCD detector columns, point spread function correction is
% performed using \textmd{xrtlccorr}. The light curve uses a bin
% size of one observation per bin (observations were typically of 1
% sec duration).
%GG%Each observation was fit spectrally using a photon binning ratio that ensured more than 20 photon counts per energy bin.
%GG%This typically resulted in more than 55 energy bins in the spectrum and dropped down to a minumum of 20 energy bins per observation
%GG%for the smallest snapshots.
%GG%, which is more than sufficient to contrain the spectral parameters.
%GG%Once the spectral fits were obtained,
%GG%the flux was calculated in the 0.3-10 keV range for the light curve.\\
%
We fitted the data with an absorbed power-law model.
We obtained photon indices between
1.51 $\pm$ 0.09 and 1.73 $\pm$ 0.11,
and excess absorption between (0.09 $\pm$ 0.06)$\times 10^{22}$ and (0.17 $\pm$ 0.03)$\times 10^{22}$ cm$^{-2}$ (all uncertainties on xrt spectral fit are at 90\% level).
%
%GG%We fitted the data with an absorbed power-law model, fixing the absorption coefficient to 0.13 $\times 10^{22}$ cm$^{-2}$.
%GG%We obtained photon indices between 1.51 $\pm$ 0.11 and 1.73 $\pm$ 0.10.
%GG%and an excess absorption between (0.09 $\pm$ 0.06)$\times 10^{22}$ and (0.17 $\pm$ 0.06)$\times 10^{22}$ cm$^{-2}$.
% \subsection{optical data}
%

% \subsection{\emph{Swift}/UVOT}
%\textit{Swift}-UVOT data
%%GG%have been pre-processed at the \emph{Swift} Data Center and
%%GG%require only minimum user processing. The image data of each filter,
%from each observation sequence,
%%GG%i.e., with a given observation ID,
%were processed by a standard procedure.
%% summed using {\tt
%% uvotimsum}. After visual inspection, photometry of the source in
%% individual sequences are derived via {\tt uvotmaghist} which
%% ncludes a coincidence loss correction.
\textit{Swift}-UVOT data from each observation sequence were
processed by the standard UVOT tool  \texttt{uvotsource} using the
same version of the Swift software as for the XRT analysis. An
extraction region of radius 5 arcsec centered on the source and a
suitable background region were used. Magnitudes are based on the
UVOT photometric system \cite{Poole08}.
%

% \subsection{Kanata}
The optical photometry of the Kanata Observatory data was
performed using TRISPEC \cite{watanabe2005}. The observations were
pipeline-reduced, including bias removal and flat-field
corrections. We derived the $V$-band magnitude from differential
photometry with a nearby reference star, USNOB 1061-0614254
%GG%($\alpha_{2000}=22^{\rm h}53^{\rm m}58^{\rm s}.1$,
%GG%$\delta_{2000}=+16^{\circ}09'07''$, $V=13.587$,
\citep[
%GGG%$\alpha_{2000}=22^{\rm h}53^{\rm m}58^{\rm s}.1$,
%GGG%$\delta_{2000}=+16^{\circ}09'07''$, $V=13.587$,
see][]{gonzales2001}. Photometric
%GG%invariability
stability of this star has been confirmed by our simultaneous photometry for another nearby
star, USNOB 1061-0614207.
All REM raw optical and NIR frames, obtained with ROSS
\cite{tosti2004} and REMIR \cite{conconi2004} respectively, 
as well as images from GRT
were corrected for dark, bias, and flat field following standard
recipes. Instrumental magnitudes were obtained via aperture
photometry, and absolute calibration has been performed by means
of secondary standard stars in the field \cite{raiteri1998}.
%

% \subsection{Radio observations}
Even if it is not reported in Figure~1, the
Mets\"ahovi radio data  at 37 GHz show a high flux with an
increasing trend from 2009 December 1 until 2010 January 14, and a
mean flux of $\sim$ 20 Jy during the first week of December, 2009.
The mean 230 GHz flux is $\sim$ 25 Jy.
%GGGG%We note that these measured
These
radio fluxes\footnote{The long term radio and optical light curves
of \c during the 2009-2010 observing season will be presented
in a forthcoming paper (Raiteri et al., in preparation).} are
comparable to the peak flux measured in 2006 \cite{villata2007}.

All UV/optical/NIR data presented here were corrected for the Galactic extinction toward 3C~454.3
assuming $A_{V}=0.349$ \cite{schlegel1998}.

\section{Results}
% \subsection{The light curves}
%\begin{figure*}
\begin{figure}
\centerline{\psfig{figure=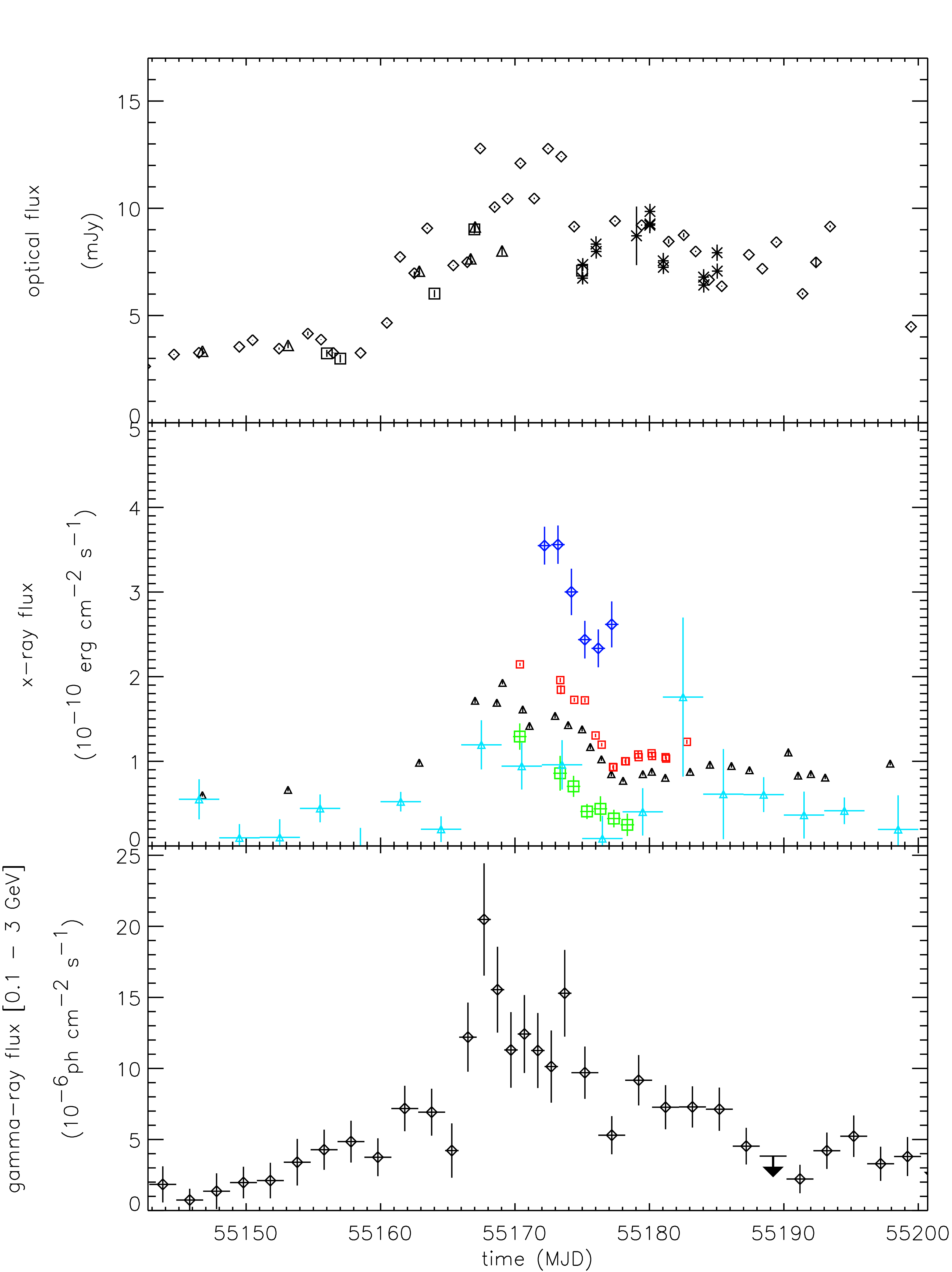,width=94.mm,height=130.mm}}
%\centerline{\psfig{figure=3c454.3_lc_all_page4_bat_original_remfiltered.ps,width=94.mm,height=130.mm}}
%\centerline{\psfig{figure=3c454.3_lc_all_page4_bat_original.ps,width=94.mm,height=130.mm}}
%%\centerline{\psfig{figure=3c454.3_lc_all_page4.ps,width=94.mm,height=130.mm}}
%%%\centerline{\psfig{figure=3c454.3_lc_all_page3.ps,width=80.mm,height=130.mm}}
%%%%\centerline{\psfig{figure=3c454.3_lc_all_page3.ps,width=180.mm,height=230.mm}}
%%%%%\psfig{figure=3c454.3_lc_all_page3.ps,width=180.mm,height=230.mm}
\caption{Multifrequency data collected during the
%GGG%AGILE observations of 3C 454.3.
campaign
\textit{(Upper panel:)}
%GGG%optical V data corrected for galactic extinction.
{\rffa{Optical V data from: KANATA (diamonds), REM (star), GRT
(squares), UVOT (triangles).}}
\textit{(Middle panel:)} X-ray data
from XRT (0.3-10 keV flux, black triangles), PCA (3-20 keV, red
squares), HEXTE (20-40 keV, green squares), BAT (14-195 keV rate,
rescaled to the 20-40 keV band, cyan triangles), ISGRI (20-200 keV, blue
diamonds). \textit{(Bottom panel:)} gamma-ray data from the
AGILE-GRID (0.1-3 GeV, black diamonds).}
\label{fig:lc_all}
%\end{figure*}
\end{figure}
%
%GG%\begin{figure*}
%GG%\centerline{\psfig{figure=3c454.3_lc_all_magn.ps,width=180.mm,height=120.mm}}
%GG%\caption{light curves of 3C 454.3 for V band data (black, measured magnitude units), X-ray from XRT (red, in units of 2.5$\times$log(flux[erg cm$^{-2}$ s$^{-1}$])-9.9 ), hard X ray from BAT (cyan, 2.5$\times$log(flux[mCrab])+17.5),
%GG%and gamma-ray from AGILE (green, 2.5$\times$log(flux[10$^{-6}$ph cm$^{-2}$ s$^{-1}$]+16))}
%GG%\label{fig:lc_all_magn}
%GG%\end{figure*}
%
The multifrequency light curves of \c are reported in Figure
\ref{fig:lc_all}. The exceptional gamma-ray flaring activity
%GGG% of \c
is  produced during an extended period lasting several weeks.
{\rffa{The broad-band coverage is excellent,
and for the first time allows a detailed multi-wavelength study of
a gamma-ray blazar peaking above the Vela pulsar flux  \citep[for
comparison, see the multifrequency coverage for the
gamma-ray super-flare of PKS~1622-29,][]{mattox1997}.}}

%GGG%
%GGG%The broad-band coverage of the gamma-ray activity is excellent,
%GGG%and for the first time allows a detailed multi-wavelength study of
%GGG%a gamma-ray blazar peaking above the Vela pulsar flux  \citep[for
%GGG%comparison, see the multifrequency coverage obtained for the other
%GGG%gamma-ray super-flare known to date from a blazar, i.e., that of PKS~1622-29,][]{mattox1997}.
%GGG%

%
The optical data show variability timescale as short as one day or
less. Due to the typical interval between optical observations (1 day), we did not
show in Fig. \ref{fig:lc_all} a detailed representation of the
variability. The optical V data indicate a low-state of the source
(V $>$ 15 mag, flux $<$ 3.9 mJy) until MJD 55160, after which the
flux started to increase. The optical flux increased of
about 50\% in less then one day from MJD 55166.4 until MJD
55167.4, reaching the value $V=13.7$ mag (e.g., 12.8 mJy),
followed by a fast flux decrease to  $V=14.3$ mag (e.g., 7.4 mJy)
at MJD 55169.0. Another optical peak was reached at MJD 55172.5
(V=13.7 mag), and then was followed by a minimum at MJD 55175.0,
with V=14.4 mag (e.g., 6.7 mJy).

In general, the X-ray flux follows\footnote{We note that no X-ray
data were obtained in exact correspondence with the gamma-ray
super-flare of Dec. 2-3, 2009.} the rising part of the optical
emission in the interval MJD 55150-55169. Starting on MJD~55169,
the X-ray flux started to fade, with the optical emission
remaining in a relatively high state
% light curve show a variable flux but comparable to the optical peak
for 4-5 days.
% optical maximum at MJD 55167.4.
 Our INTEGRAL hard X-ray data sample the
fading phase of the high-activity period.

\begin{figure*}
\centering
\begin{tabular}{c}
\psfig{figure=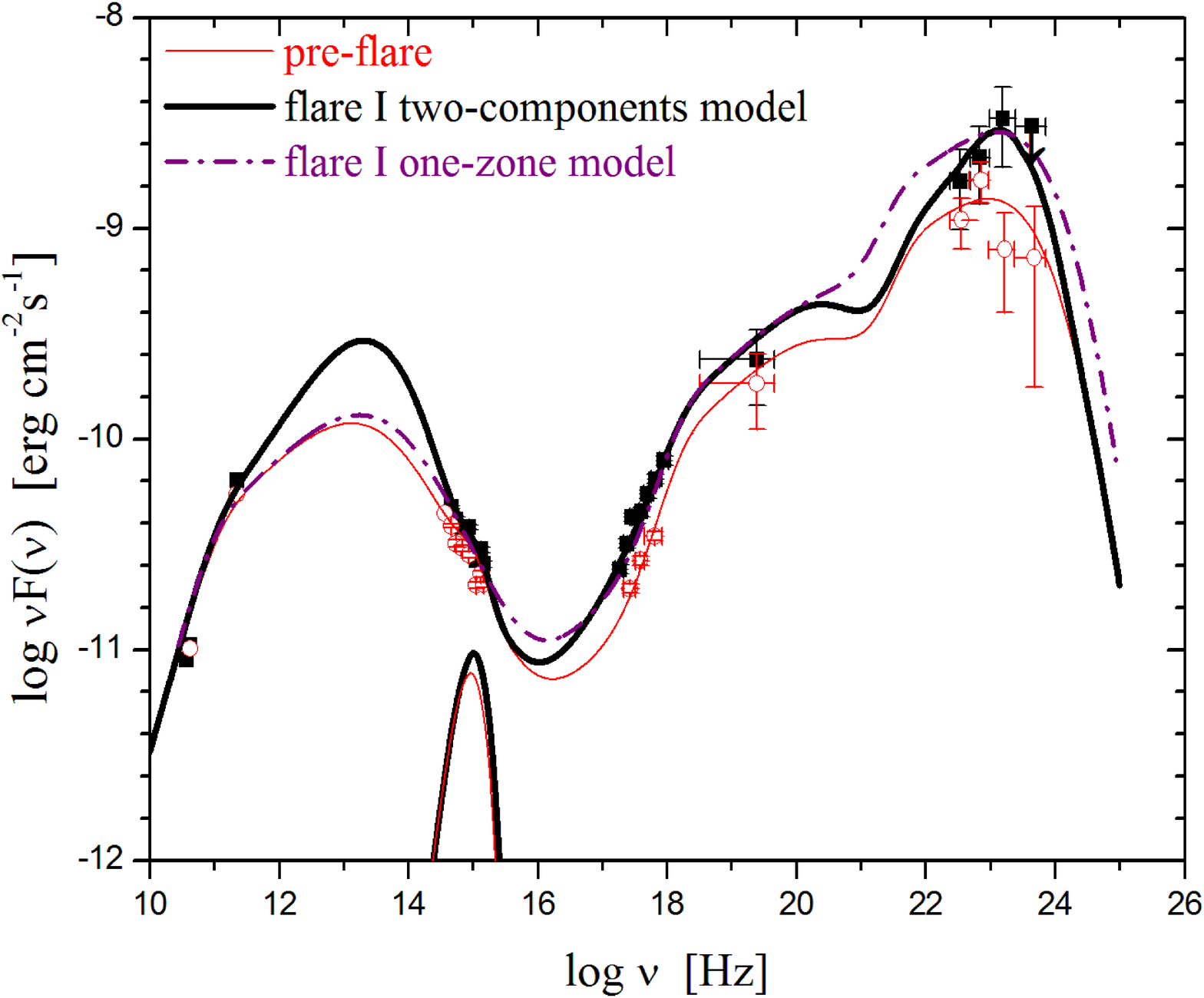,width=122.mm,height=80.mm}  \\ \psfig{figure=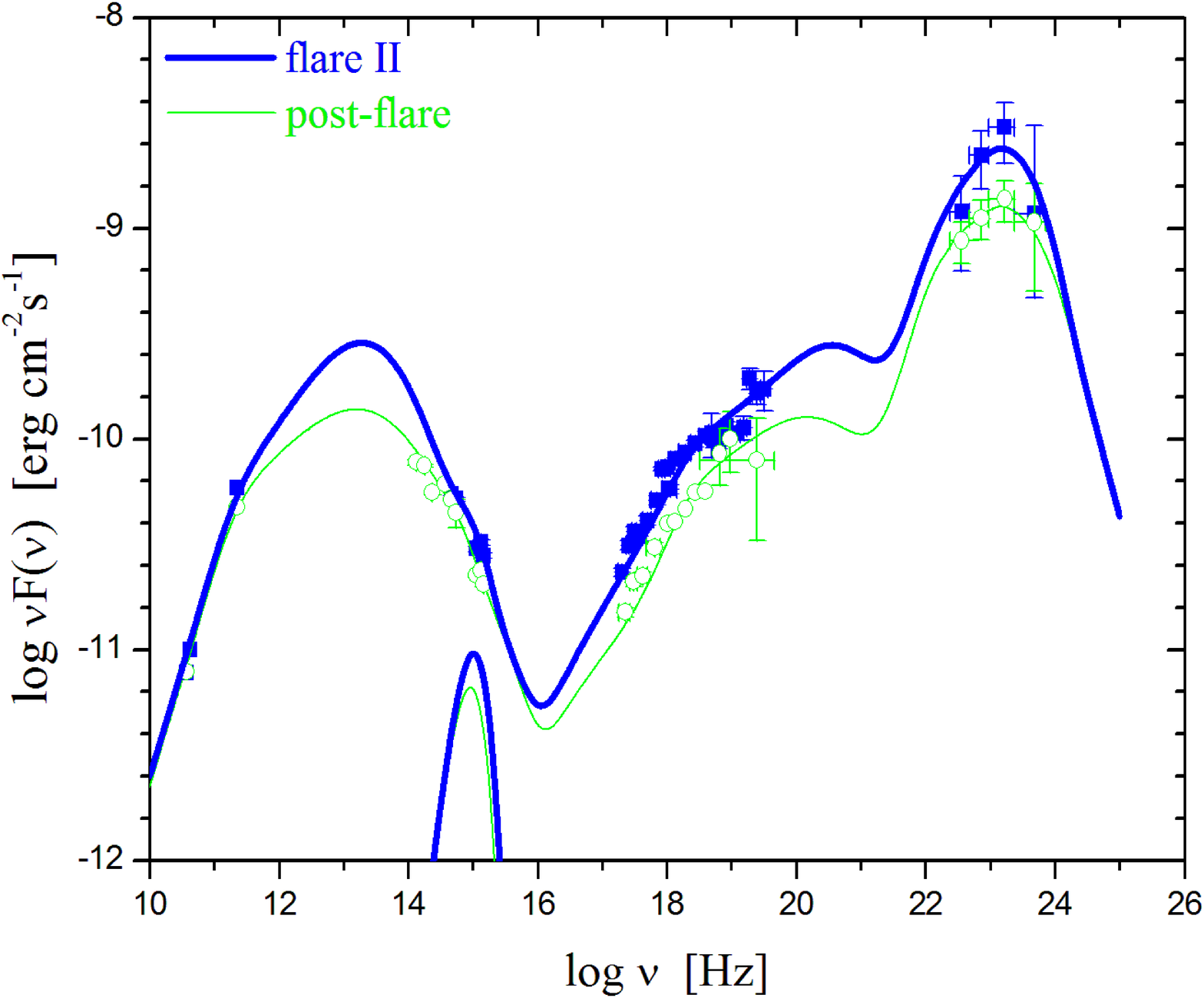,width=122.mm,height=80.mm}
%%GG%\psfig{figure=superflare_left_v5_mod.eps,width=92.mm,height=65.mm}  & \psfig{figure=superflare_right_v5_mod.eps,width=80.mm,height=65.mm}
\end{tabular}
\caption{ Multifrequency simultaneous spectra of \c at four
different epochs superimposed with spectral modelling. {\rffa{Top}}
panel: pre-flare (red open circles) and super-flare (black filled squares). {\rffa{Bottom}} panel: secondary
flare (blue filled squares) and post-flare spectra (green open circles).
Models details are explained in the text.
%GG%{\rffa{The models for the pre- and post-flares are reported as thin solid lines.
%GG%The two-component models for the super- and secondary-flare are reported as thick solid lines. 
%GG%The one-zone model for the super-flare is reported as dot-dashed line.}}
}
\label{fig:sed}
\vskip 0.3in
\end{figure*}

{\rffa{We performed a Discrete Correlation Function (DCF) analysis for the optical-gamma, and soft x-gamma data set.
No significant delay is found within 2 days (compatible with the average bin of gamma-ray light curve).}}

We focused on the time-dependent spectral analysis of this
exceptional activity of \cp, and used simultaneous broad-band data
to obtain a detailed account of the source variability. We
obtained the SED for four periods.
The first period (interval-1) is for a 5-day integration
of GRID data, centered on 2009-11-27 09:36 UT (MJD 55162.4), and
using the \emph{Swift} observation at MJD 55162.9, and
quasi-simultaneous GASP optical data (from KVA and New Mexico
Skies) obtained within 16 hours from the \emph{Swift} observation
(pre-flare SED).
%GG%Roque (KVA) observations at 2009-11-27 20:44 UT, New Mexico Skies at  2009-11-28 04:43, at 2009-11-29 00:13 UT (pre flare SED).

The second period (interval-2) was
obtained for the gamma-ray super-flare episode
integrating the  GRID data for 1 day centered at
2009-12-02 16:48 UT (MJD 55167.7), and using of
the simultaneous \emph{Swift} and GRT observations at MJD 55167.0
(first flare SED).
%GG%(we notice that these optical and soft X-ray data were taken within approximately 2 hours from each other).
% at the peak of the optical
% flare, but with this choice, X-ray and optical data are
% simultaneous within 2 hours.
The Kanata observatory measured a flux 40\% higher in
V band 10 hours after the GRT observation.

The third SED was obtained (interval-3) integrating GRID
data for 2 days centered at 2009-12-06 16:48 UT (MJD 55172.7), to
match the local maximum of the optical light curve with
V=13.7 mag, that is apparently coincident with the secondary
gamma-ray maximum near MJD~55174. For this interval we used the
INTEGRAL-ISGRI data collected between MJD 55171.7 and MJD 55174.2,
\emph{Swift} data at MJD 55173.9, RossiXTE data at 55173.4,
and GASP optical data (from Lulin and GRT) obtained  within 26 hours
from the \emph{Swift} data.
%GG% Lulin (SLT) observations at 2009-12-09 10:53, and GRT at 2009-12-09 23:43 UT (second flare SED).

The last SED (interval-4) was obtained integrating GRID
data for 5.5 days centered at 2009-12-15 18:00 UT (MJD 55180.8),
and making use of the \emph{Swift} observation at MJD 55179.1, the
RossiXTE observation at MJD 55179.2, the GASP optical data (from
St. Petersburg and Lulin) simultaneous with \emph{Swift} within 12
hours,
and the near-infrared observations from REM at MJD 55181.0 (post flare SED).\\
Radio data for the SEDs were taken by the GASP,
using observations from Mauna Kea (SMA, 230 GHz), Noto (43 GHz), Mets\"ahovi (37 GHz)
UMRAO (4.8, 8.0, 14.5 GHz), simultaneous within a few days with the
XRT observations. Due to the slow variability of the radio data,
we use interpolated radio values in the SEDs.
The results are shown in Fig.~\ref{fig:sed}.\\

\section{Discussion}

The multifrequency data of the extensive campaign on \c show a
remarkable behavior of the source.
% During November-December 2009 the source was observed in a quite
% active phase, with a \gray flux exceeding 200 \phcmsec and around
% a magnitude 14 in V band. In detail, s
Starting from MJD 55150, first an overall rise of the gamma-ray
emission  and then of the X-ray and optical fluxes is detected.
This rise culminates with peak optical/X-ray/gamma-ray
emission during a 10-day period centered around MJD 55173.
Subsequently, the overall flux decreased and reached again a
relative-minimum state around MJD 55200. During the
2-month period the optical and X-ray fluxes vary within a factor
of 3, whereas the \gray flux grows by a factor of 5-10 compared to
the pre-flare value. During the rapid super-flare around MJD~55167.7
the \gray flux doubles within 1 day with the optical and
average X-ray  increase of 50\% and 30\%, respectively.
% only and with an X-ray flux increased of 30$\%$ with respect to three days before.
%GGG%A similar behavior is detected $\sim6$ days later during the
%GGG%secondary \gray flare on MJD~55172.7
% We briefly emphasize here the most relevant facts concerning our observations.

We find an overall correlation at all wavelengths for both
long and short timescales. However, the unusual gamma-ray flaring
and super-flaring activity from \c during the period
November-December, 2009 is not accompanied by strong emission of
similar intensity in the optical or even in the soft X-ray
bands. {\rffa{The hard x-ray flux
is comparable to the level of spring 2005, e.g., $\sim 3.5\times 10 ^{-10}\ erg\ cm^2\ s^{-1}$ in 20-200 keV \cite{pian2006}.}}
This flaring behavior appears to be quite different from
other episodes detected in 2007 and 2008
\citep[e.g.][]{vercellone2009a,donnarumma2009a}.
% \textbf{In
% particular, the observed optical variations that accompany the
% gamma-ray peaks are not so strong as the \gray ones. This strongly
% constrains the theoretical models.}
The synchrotron emission appears to be quite broad and centered
around $\nu \sim 10^{13}$~Hz.

\begin{table*}
\begin{center}
\caption{
Models parameters used for producing the spectral analysis of Fig. 2.
%GG%{\rff Models} parameters used for producing the {\rff two-component} spectral analysis of Fig. 2 (filled lines).
The columns give the emission
interval specification, the leptonic component
that dominates the gamma-ray emission, the
magnetic field $B$, the comoving emission radius
$R$, the particle energy distribution
normalization $K$, the break energy (Lorentz
factor) $\gamma_b$, the minimum particle energy
$\gamma_{min}$, and the low- ($\zeta_1$) and
high-energy spectral indices ($\zeta_2$) defined as
%  $N(\gamma) = K \,\gamma_b^{-1} (\gamma/\gamma_b)^{-\zeta}$.
$N(\gamma) = K \,\gamma_b^{-1} / [(\gamma/\gamma_b)^{\zeta_1} +
(\gamma/\gamma_b)^{\zeta_2}] $. The relativistic Maxwellian is
defined as $N(\gamma) = K \,(\gamma/\gamma_b)
\exp(-\gamma/\gamma_b)$.}
\label{table:sed-bi-zone}
 \small \noindent
\scriptsize
\begin{tabular}{|l|c|c|c|c|c|c|c|c|c|c|}
  \hline
  \bf{Interval}       & \bf{Model}    &\bf{Component} &  \bf{B} & \bf{R} & $\bf{K}$    & $\bf{\gamma_b}$ & $\bf{\gamma_{min}}$ & $\bf{\zeta_1}$ & $\bf{\zeta_2}$ & Comments \\
                      &               &                          &  (G)    & (cm)   & $(cm^{-3})$ &                 &                     &                &                &          \\
  \hline              
  1 (pre-flare)       & two-comp.     & component-1$^*$   &  0.6 & $7\times10^{16}$ &2.2   & 800 & 35 & 2.35 &4.5 & broken PL\\ \cline{3-11}
                      &               & component-2   &  --  & --              & --   & --  & -- & --   & -- & --        \\ \hline
  2 (super-flare)     & two-comp.     & component-1   &  0.6 & $7\times10^{16}$ &2.2   & 800 & 35 & 2.35 &4.5 & broken PL\\ \cline{3-11}
                      &               & component-2   &  0.9 & $3\times10^{16}$ &180   & 180 & 1  &  --  & -- & relativistic Maxwellian  \\  \cline{2-11}
                      &\multicolumn{2}{|c|}{one-zone} &  0.5 & $7\times10^{16}$ &2.5   &1000 & 35 & 2.35 & 4.5& broken PL\\ \hline
  3 (secondary-flare) & two-comp.     & component-1   &  0.6 & $7\times10^{16}$ &2.5   & 800 & 45 &2.25 &4.5 & broken PL\\ \cline{3-11}
                      &               & component-2   &  0.9 & $3\times10^{16}$ &170   & 170 & 1  & --  & -- & relativistic Maxwellian \\ \hline
  4 (post-flare)      & two-comp.     & component-1$^*$   &  0.6 & $7\times10^{16}$ &2.5   & 800 & 45 &2.25 &4.5 & broken PL\\ \cline{3-11}
                      &               & component-2   &  --  & --              & --   & --  & -- & --   & -- & --       \\ 
  \hline
\end{tabular}
\end{center}
\normalsize
$(^*)$For the pre- and post-flare intervals this set of parameters describes also the simple one-zone model.
\end{table*}
%GG%\end{table*}
%\normalsize

Striani et al. (2010)
provided a first report of the AGILE-GRID data.
%GGG%(optimized {\lp in the 100 MeV - 1 GeV band}).
%GG%at energies near 100 MeV)
%
%GG%indicate that the super-flare
%GG%of Dec. 2-3, 2009 indicate a possible spectral hardening.
%GG%A single power-law approximation gives average photon spectral
%GG%indices in the energy band 0.1-1 GeV ranging from $1.9\pm
%GG%0.1$ to $1.6\pm 0.3$ within 2 days.
%
A single power-law approximation gives a photon spectral
index 1.66 $\pm$ 0.32 in the energy band 0.1-1 GeV, integrating the
data for two days centered at MJD 55167.7.
%
%GG% ranging from $1.9\pm 0.1$ to  within 2 days.
We confirm this result in the analysis presented
here integrating gamma-ray data for 1 day,
as shown in our Fig. 2 (left panel). In addition, we found
a similar spectral shape of the gamma-ray emission during
the secondary maximum of interval-3. Our Fig. 2 (right panel)
shows the interval-3 SED
%GGG%(for which the INTEGRAL-IBIS data provide crucial information)
as compared with the post-flare SED of
interval-4.

The simultaneous observations of \c by AGILE, INTEGRAL, RossiXTE and
\textit{Swift} strongly constrain the emission models in the
high-energy range. In particular, our interval-3 spectrum (blue solid
squares of Fig. 2, right panel) represents one of the best
constrained multifrequency spectra ever obtained for a flaring
blazar from X-ray up to GeV energies.

We present in Fig. 2 the results of our spectral modelling based on
{\rffa{Synchrotron-Self}} Compton (SSC),
%GG%synchrotron, SSC,
{\rffa{plus contribution from external seed photons (EC).}}
%GGG%inverse Compton emission.
We used  parameters similar
to those already implemented to model previous gamma-ray flares of
3C~454.3.
%GG%that used as input model parameters similar to those already implemented to
%GG%model other gamma-ray flares of \c
\cite{vercellone2009a,donnarumma2009a}.
We find that the pre-
and post-flare spectra (interval-1 and interval-4) are adequately
represented by a simple one-zone SSC model plus EC
%GG%of emission
in which the accretion
disk and the broad-line region provide the necessary soft
radiation field for the inverse Compton components that dominate
the X-ray through the GeV energies.
{\rffa{The simple one-zone model parameters for the pre- and post-flare intervals are reported in Table \ref{table:sed-bi-zone} (they are also labelled \emph{component-1} in the table).}}
%GG%\begin{table}
%GG%\begin{center}
%GG%\caption{One-zone model parameters for the 2009 \gray super-flare.}
%GG%\label{table:sed-one-zone}
%GG% \small \noindent
%GG%\scriptsize
%GG%\begin{tabular}{|c|c|c|c|c|c|c|l|}
%GG%  % after \\: \hline or \cline{col1-col2} \cline{col3-col4} ...
%GG%  \hline
%GG%  \bf{B} & \bf{R}                & $\bf{K}$    & $\bf{\gamma_b}$ & $\bf{\gamma_{min}}$ & $\bf{\zeta_1}$ & $\bf{\zeta_2}$ & Comments \\
%GG%     (G) & (cm)                  & $(cm^{-3})$  &                 &                    &                &                &          \\ \hline
%GG%   0.5   & $7\times10^{16}$       &      2.5    &         1000    &       35           &     2.35       & 4.5            & broken PL\\
%GG%  \hline
%GG%\end{tabular}
%GG%\normalsize
%GG%\end{center}
%GG%\end{table}
%GG%

{\rffa{We can fit the super-flare with a simple one-zone model as shown in Figure \ref{fig:sed} (dot-dashed curve), with the parameters reported in 
Table \ref{table:sed-bi-zone}.
Such a model requires (compared to the pre-flare fit parameters) an increase of the electron energy and density, and a slight reduction of the comoving magnetic field
for the whole  electron population of the blob  \citep[as also reported in ][]{bonnoli2010}.}}
{\rffa{We note, however, that we used the optical data obtained during the rising edge of optical emission for the SED of the super-flare.
Using only the V peak emission for the optical portion of the spectra, the one-zone model gives a magnetic field of 0.55 G, closer to the value for the pre-flare.}}

{\rffa{The time evolution of the source (e.g., the lack of strong optical emission and
the gamma-ray spectrum during the super-flare) can be also described adopting a different approach.}}
%
%GGG%However, from the lack of simultaneous strong optical (and X-ray)
%GGG%emission we find that a simple one-zone model is
%GGG%problematic to explain  the super-flare (interval-2) and secondary
%GGG%flare episodes (interval-3).
%
%GGG%We briefly report here the results of our extensive calculations.
We assume a long-term rise and fall of the mass accretion rate
onto the central black hole. This enhanced accretion causes an
overall increase of the synchrotron {\rffa{emission}} and of the soft
photon background scattered off by the primary component of
accelerated electrons (component-1).
An additional population of accelerated leptons (component-2,
co-existing with component-1) {\rffa{can be}} introduced for the super-flare
and secondary flare episodes. This component is a consequence of
additional particle acceleration and/or plasmoid ejection near the
jet basis.
Table~1 reports the parameters that we used to model the Fig.~2
spectra {(the two-component models are reported as {\rffa{thick}} solid lines for the super-flare and for the secondary flare).}
{\rffa{For all intervals, we assume the presence of
component-1. This is the only component acting for the pre- and post-flare intervals, with the parameters given in Table~1.
The super-flare (secondary-flare) parameters for the component-1 are the same as the pre-flare (post-flare) interval.}}
We assumed a bulk Lorentz factor
$\Gamma = 25$, a jet angle with respect with the line of sight
$\theta=1.2^o$, an accretion disk
% illuminating the jet with a
of bolometric luminosity $L_d=6 \times
10^{46}$\ergsec slowly decaying toward $L_d= 5
\times 10^{46}$\ergsec. A broad line region
located 0.5 pc  from the black hole reflects 5\%
of the disk power toward the emitting regions.

The component-2 energy distribution that
better reproduces our gamma-ray spectral data is
a relativistic Maxwellian of characteristic
energy $\gamma_b \simeq 180$. Interestingly, this
component appears to be strongly energized but
not yet modified by additional non-thermal
acceleration.

{\rffa{This two-component approach avoids the problem of explaining the
time variability of the physical parameters of the whole electron population of the one-zone model, at the price of
adding an additional varying component superimposed with the pre-existing one.}}

To summarize, our multifrequency data for the December, 2009 flare
of \c provide a wealth of very important information on this
puzzling and fascinating blazar. We find that \c is characterized
by strong broad-band spectral variability, and that the modelling
of the peak gamma-ray emission episodes
%GG%require a modification of standard modelling.
{\rffa{suggests more elaborate models than the standard one-zone SSC+EC models
of bright blazars.}}

\bigskip
% \section*{Acknowledgments}

The AGILE Mission is funded by the Italian Space Agency (ASI) with
scientific and programmatic participation by the Italian Institute
of Astrophysics (INAF), and the Italian Institute of Nuclear
Physics (INFN). This investigation was carried out with partial
support from the ASI contract n.~I/089/06/2. V.Larionov
acknowledges support from Russian RFBR foundation via grant
09-02-00092. The operation of UMRAO is made possible by funding
from the NSF, NASA, and the University of Michigan. The
Submillimeter Array is funded by the Smithsonian Institution and
the Academia Sinica Institute of Astronomy and Astrophysics.
The GASP president acknowledges the ASI support through
contract ASI-INAF I/088/06/0.
%
%\section*{References}
%\begin{harvard}

%\end{harvard}

\end{document}